\documentclass{moriond}

\def\Journal#1#2#3#4{{#1} {\bf #2}, #3 (#4)}

\def\be{\begin{equation}}
\def\ee{\end{equation}}
\def\bea{\begin{eqnarray}}
\def\eea{\end{eqnarray}}

\newcommand{\Photo}{}

\begin{document}
\begin{flushright}
IFT-UAM/CSIC-18-046  \\
FTUAM-18-11\\
LPT-Orsay-18-64
\end{flushright}

\vspace*{1cm}
\title{RESONANT $pp\to W^+ Zjj$ EVENTS AT THE LHC \\{\footnotesize FROM A  UNITARIZED STUDY OF THE ECHL}}

\author{ R. L. DELGADO$^1$, A. DOBADO$^2$, D. ESPRIU$^3$, C. GARCIA-GARCIA$^4$\footnote{Talk given by C.G.G. at the $53^{\rm rd}$ Rencontres de Moriond on Electroweak Interactions and Unified Theories.}, M. J. HERRERO$^4$,\\ X. MARCANO$^5$, J. J. SANZ-CILLERO$^2$}

\address{
\scalebox{0.8}{
$^1$Physik-Department T30f, Technische Universitat Munchen, James-Franck-Str. 1, D-85747 Garching, Germany.}\\
\scalebox{0.8}{
$^2$Departamento de F\'isica Te\'orica and UPARCOS, Universidad Complutense de Madrid,
Plaza de las Ciencias 1, 28040 Madrid, Spain.}\\
\scalebox{0.8}{
$^3$Departament de F\'isica Qu\`antica i Astrof\'isica and Institut de Ci\`encies del Cosmos (ICCUB),
Universitat de Barcelona,}\\\scalebox{0.8}{Mart\'i i Franqu\`es 1, 08028 Barcelona, Catalonia, Spain.}\\
\scalebox{0.8}{
$^4$Departamento de F\'{\i}sica Te\'orica and Instituto de F\'{\i}sica Te\'orica, IFT-UAM/CSIC,
Universidad Aut\'onoma de Madrid,}\\\scalebox{0.8}{ Cantoblanco, 28049 Madrid, Spain.}\\
\scalebox{0.8}{
$^5$Laboratoire de Physique Th\'eorique, CNRS,
Univ. Paris-Sud, Universit\'e Paris-Saclay, 91405 Orsay, France
}
}

\maketitle\abstracts{
We present a study of the production of vector resonances at the LHC via $W^+Z$ vector boson scattering and explore the sensitivities to these resonances for expected LHC luminosities. We work in the framework of the electroweak chiral Lagrangian, where these resonances can be generated dynamically by unitarizing the scattering amplitudes.  
We implement all these features into a model adapted for MonteCarlo, the IAM-MC, that allows us to give predictions for the sensitivity to these resonances and to the relevant parameters involved for $pp \to W^+Zjj$, $pp\to \ell_1^+\ell_1^-\ell_2^+\nu jj$, and $pp \to JJjj$.
}
\section{Introduction}
The Higgs boson was found at the LHC completing the Standard Model (SM), but its discovery posed a lot of new questions, such as if the electroweak symmetry breaking sector (EWSBS) is really as described in the SM. The answers are not yet clear, but one possibility is that the Higgs boson is a composite state, instead of a fundamental particle.

Composite Higgs models are characterized by the existence of a scale much above the EW scale where some new strong interactions trigger the dynamical breaking of a global symmetry. The Goldstone bosons that appear provide the longitudinal degrees of freedom of the weak gauge bosons, while the Higgs boson would be one of the leftover Goldstone bosons.

Along this work we will not assume an specific model for this strong dynamics, but will work instead within an effective theory approach. We will rely on the electroweak chiral Lagrangian\cite{Appelquist:1980vg,Longhitano:1980iz,Dobado:1989ax,Delgado:2014jda} with a light Higgs\cite{Alonso:2012px,Espriu:2012ih,Buchalla:2013rka} (EChL) that implements the generic and minimal assumptions for the above global group and the spontaneous symmetry breaking pattern given by $SU(2)_L \times SU(2)_R \to SU(2)_{L+R}$ as well as the same EW gauge symmetries as the SM.

Strongly interacting dynamics lead typically to the appearance of resonances in the spectrum. In the present work, these resonances, of masses naturally at the TeV scale, are generated dynamically from the EChL. We use the inverse amplitud method (IAM) to impose the unitarity of the EW gauge boson scattering amplitudes, leading to the generation of resonances that might emerge in vector boson scattering (VBS). The disadvantage of this method is that it relies on a partial waves analysis, which makes very difficult its implementation in a MonteCarlo.

In this work, we develop a model, the IAM-MC, that mimics the IAM vector resonances and that is suitable for MonteCarlo analysis. With this tool, we study the sensitivity of the LHC to these resonances when they are produced via VBS. In particular, we focus on charged vector resonances that mediate $WZ\to WZ$ scattering. We perform a careful analysis of signal and backgrounds for the full process
$pp \to WZ jj$, discussing also on the potential of the hadronic and semileptonic decays of the final $WZ$ and giving more accurate predictions for the purely leptonic channel.

All computations and details, as well as numerical results can be found in our main paper \cite{Delgado:2017cls}.
\section{The EChL and the inverse amplitud method for MonteCarlo: the IAM-MC}
The EChL is a EW gauged and chiral effective field theory coupled to a singlet scalar particle
in which the EW Goldstones bosons are introduced in a non linear representation of the global group  $SU(2)_L \times SU(2)_R$. This allows us to organize operators by their `chiral dimension' in a power momentum expansion. In this work, we include terms with chiral dimension up to  $\mathcal{O}(p^4)$ and work under the assumption that custodial symmetry is preserved. With this consideration, the relevant terms for this study\cite{Delgado:2017cls} of the leading order and the next to leading order Lagrangians read, respectively:
\begin{equation}
\mathcal{L}_2 =    \mathcal{L}_{\rm kin}^{W,Z,\gamma,H}
 +\frac{v^2}{4}\left[
  1 + 2a \frac{H}{v} + b \frac{H^2}{v^2}\right] {\rm Tr} \Big(\cal V_\mu \cal V^\mu \Big) + \dots\, ,
\label{eq.L2}\\
\end{equation}
\begin{equation}
\mathcal{L}_{4} = 
  ~ a_4 \Big[{\rm Tr}({\cal V}_\mu {\cal V}_\nu) \Big]  \Big[{\rm Tr}({\cal V}^\mu {\cal V}^\nu)\Big] 
  + a_5  \Big[{\rm Tr}({\cal V}_\mu {\cal V}^\mu)\Big]  \Big[{\rm Tr}({\cal V}_\nu {\cal V}^\nu)\Big] 
+\dots\, , \label{eq.L4}
\end{equation}
where the parameters $a,b,a_4,a_5$ are known as chiral coefficients and correspond to low energy constants encoding the information about the underlying microscopic theory.

The EW gauge boson scattering amplitudes computed within this framework typically violate unitarity. 
Because of this, we use the inverse amplitude method to make unitary predictions for these amplitudes profiting from the fact that this method is able to generate dynamically the vector resonances that we are interested in.

The properties of these resonances, such as mass $M_V$ and width $\Gamma_V$, depend on the chiral coefficients\cite{Delgado:2017cls}. In order to have resonances with masses in the TeV range, we will work with values of the chiral couplings in the experimentally allowed intervals\cite{Delgado:2017cls} $a\in[0.9,1 {\rm(SM)}]$ and $a_{4,5}\in[10^{-3},10^{-4}]$, setting $b=a^2$ as it is a well motivated relation in several models. 

Within this set of resonances we choose to explore further the isotriplet of vector resonances $V=V^0,V^\pm$. To perform our study, we select fifteen benchmark points of phenomenological interest with different values of the chiral parameters that lead to resonances with masses of 1.5, 2 and 2.5 TeV. In particular, for each mass value we select five points that differ in their value of  $a$, ranging from 0.9 to 1. 

Once we have characterized our vector resonances, we need a tool to implement them in a MonteCarlo to conduct a realistic analysis of the events of our interest.
We choose MadGraph5 and use a chiral invariant Lagrangian\cite{Delgado:2017cls,Pich:2015kwa,Ecker:1989yg} to mimic the IAM vector resonances. To obtain unitary amplitudes from this Lagrangian, we promote to a form factor the coupling between the resonance $V$ and the EW gauge bosons, $W,Z$. This way we can get unitary
 resonant amplitudes in a Lagrangian language that reproduce those of the IAM, and that can be easily  implemented in a MonteCarlo generator. We call this model IAM-MC, adapted for MonteCarlo\cite{Delgado:2017cls}.
\section{Sensitivity to vector resonances in $W^+Z$ vector boson scattering at the LHC}
\begin{figure}[t]
\begin{center}
\includegraphics[width=0.4\textwidth]{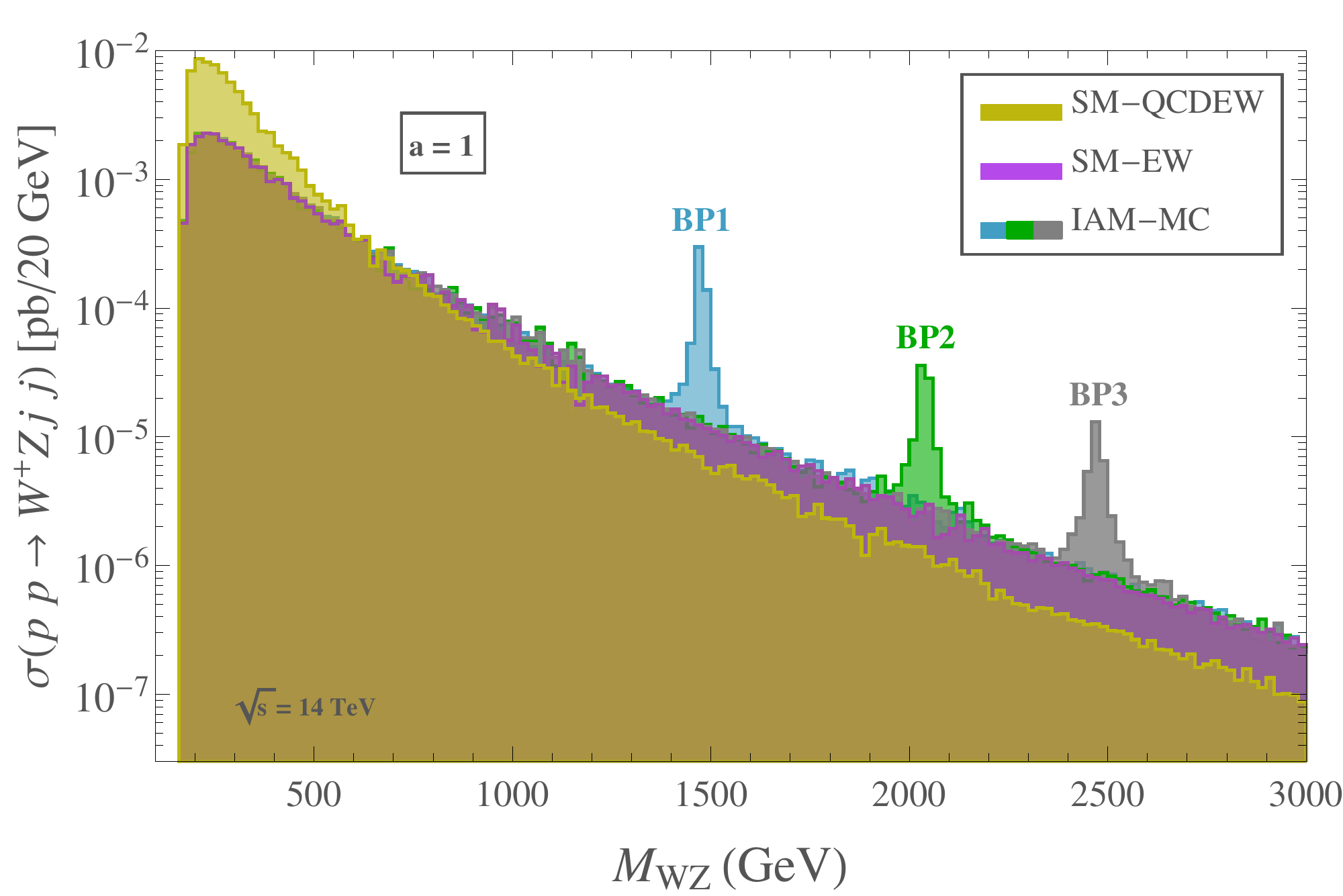}
\hspace{0.8cm}
\includegraphics[width=0.4\textwidth]{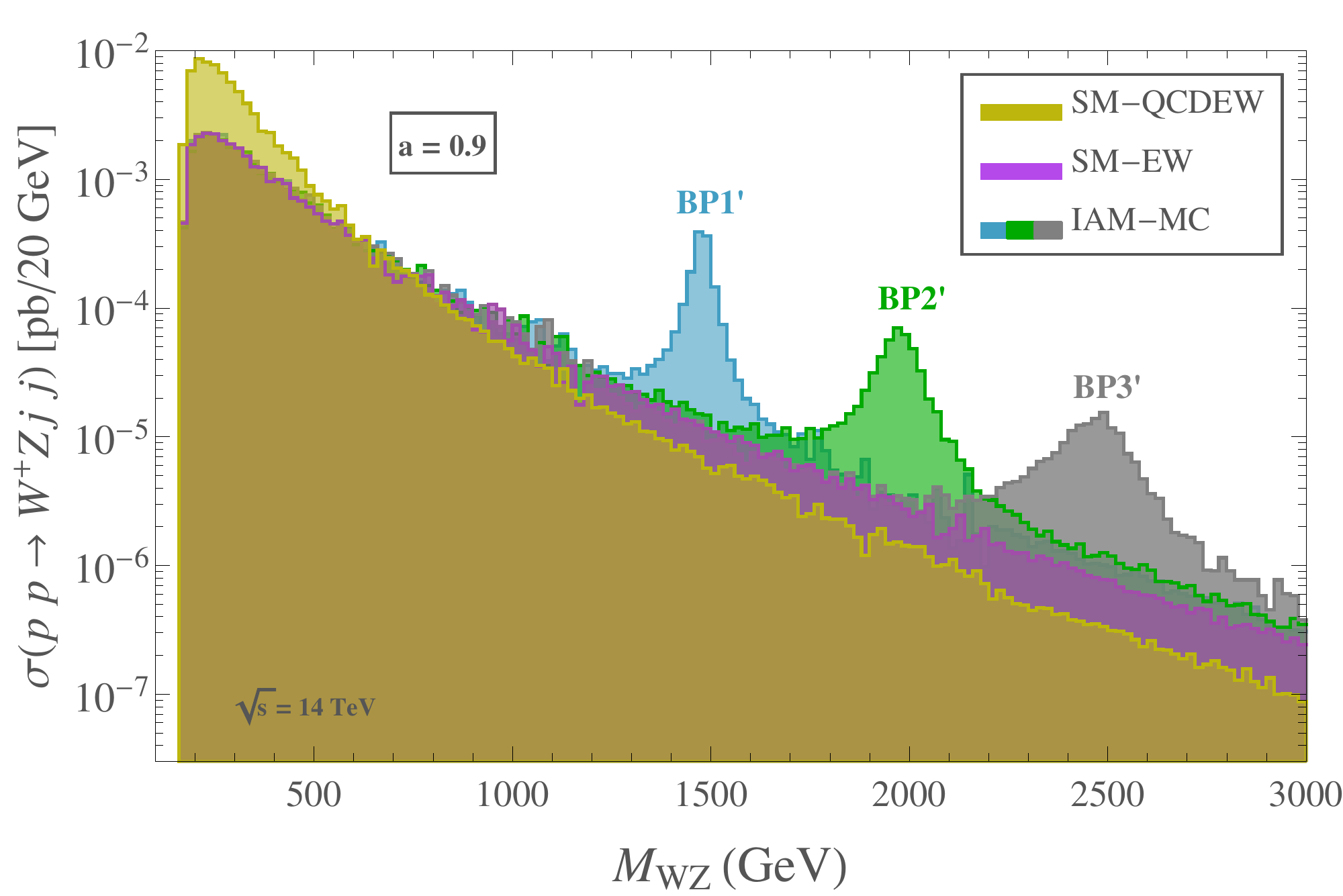}
\includegraphics[width=0.4\textwidth]{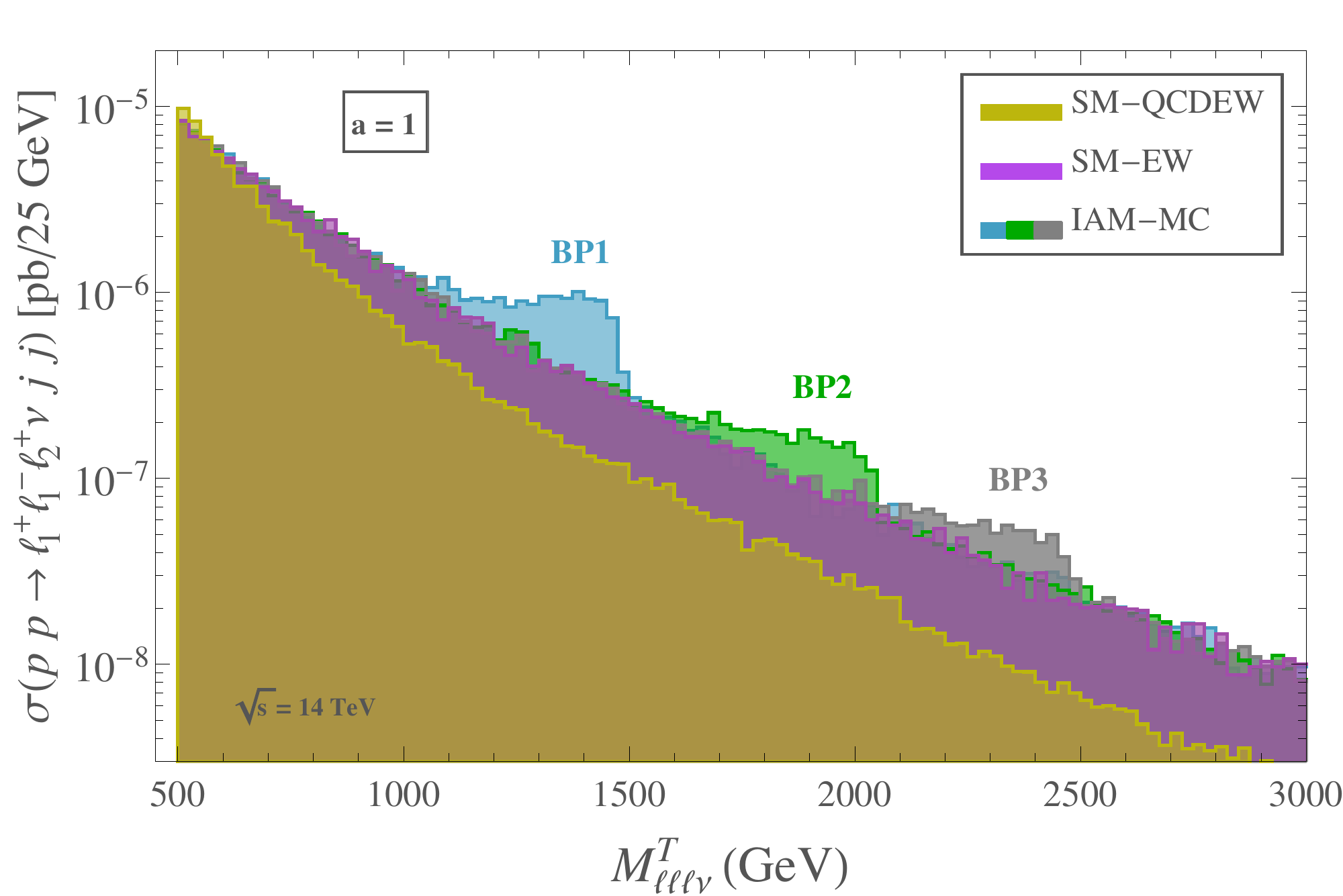}
\hspace{0.8cm}
\includegraphics[width=0.4\textwidth]{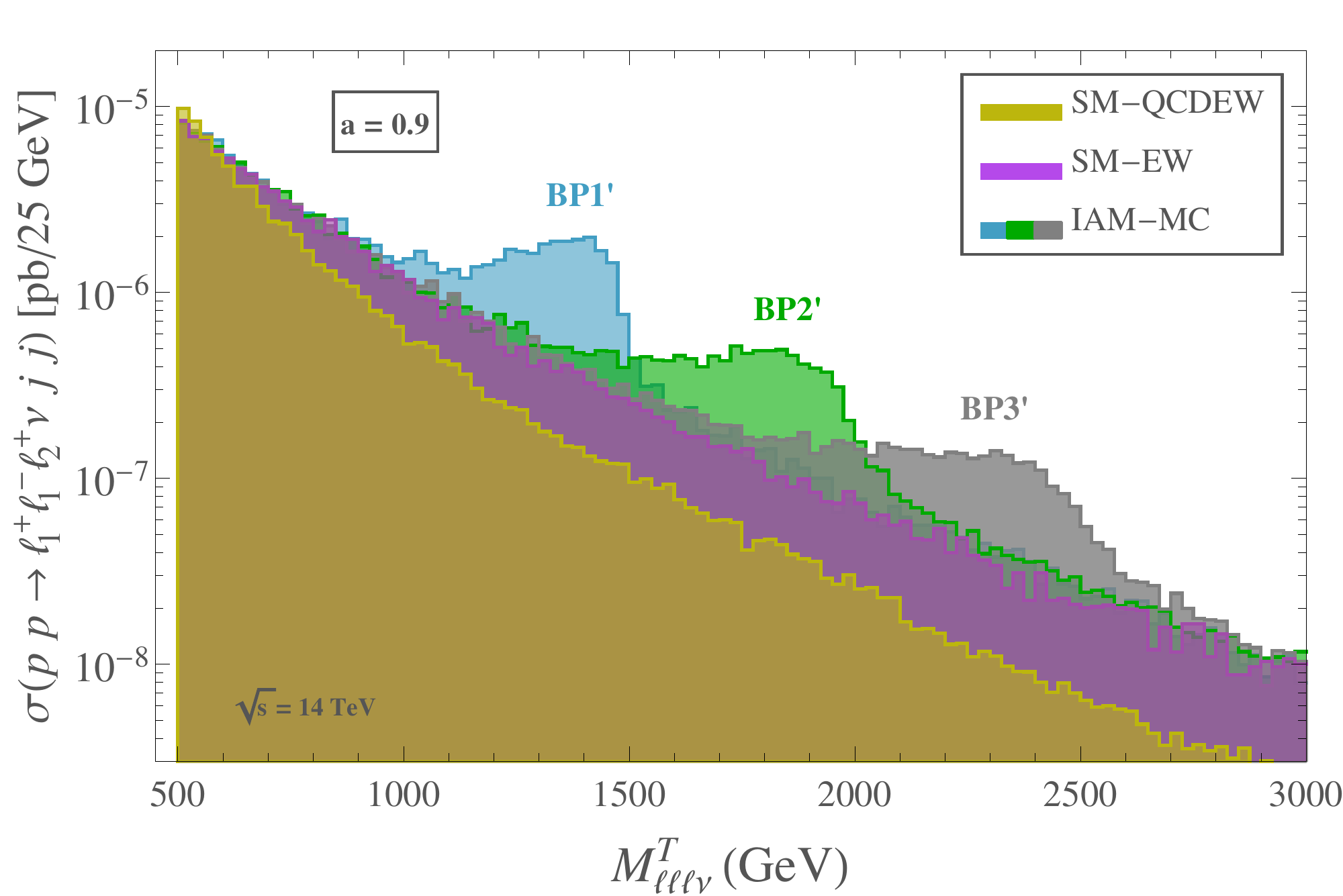}
\caption[]{Predictions of the $\sigma(pp \to W^+Zjj)$ distributions with the invariant mass of the  $WZ$ pair, $M_{WZ}$, (upper panels) and for the $\sigma(pp \to \ell_1^+\ell_1^-\ell_2^+ \nu jj)$ distributions with the transverse invariant mass, $M^T_{\ell\ell\ell \nu}$ (lower panels) for six benchmark points of the IAM-MC model\cite{Delgado:2017cls} (blue, green and gray), and of the two main SM backgrounds, SM-QCDEW (yellow) and SM-EW (purple). VBS selection cuts have been applied.}
\label{fig:WZjj}
\end{center}
\end{figure}
%
%
%

With our IAM-MC model, we simulate events to obtain predictions for $pp\to W^+Zjj$ resonant distributions. As we are interested mainly in $WZ$ scattering subprocess where $V^+$ resonates, we impose a set of cuts, based on the fact that VBS processes produce two opposite-sided large pseudorapidity jets, that allows us to select these configurations and reduce the possible backgrounds\cite{Delgado:2017cls}. We also make predictions for the two main irreducible backgrounds, that we call SM-QCDEW and SM-EW and that correspond, respectively, to amplitudes in the SM that are of order $\alpha_S\, \alpha_{EW}$ and of order $\alpha_{EW}^2$. In the upper panels of Fig.(\ref{fig:WZjj}), we display the cross section for $pp\to W^+Zjj$ resonant events vs the invariant mass of the $WZ$ system for six of the selected benchmark points, as well as for the two backgrounds. We can see that the resonances emerge clearly on top of the SM background and that there are many events available if we assume luminosities of $\sim 100 \,{\rm fb}^{-1}$. 

However, reconstructing completely the final gauge bosons is a difficult task. To make a realistic analysis one has to study events in which the $W$ and the $Z$ are detected through their decay products. This is why we also simulate the purely leptonic decay mode of the process. To this end, we impose again the VBS cuts, along with other kinematical cuts on the final state leptons, to ensure VBS selection and efficient background rejection. The cross section for $pp\to  \ell_1^+\ell_1^-\ell_2^+ \nu jj$ vs the transverse invariant mass of the four leptons for the same six benchmark points can be seen in the lower panels of Fig.(\ref{fig:WZjj}), together with the main backgrounds. It is interesting to see that some scenarios are still visible above the background, although for this case higher luminosities will be required to observe these resonances.

Finally, we perform a dedicated analysis giving predictions for the statistical significance of the observation of these resonances in the $WZjj$ case and in the $\ell_1^+\ell_1^-\ell_2^+ \nu jj$ one. We also make estimates for the process in which the gauge bosons decay to fat jets, leading to a final state of $JJjj$ scaling adequately the events for the $WZjj$ final state. The specific prescriptions for summing events and computing the significance are as described in the main paper\cite{Delgado:2017cls}.

\begin{figure}
\begin{minipage}{0.33\linewidth}
\centerline{\includegraphics[width=\linewidth]{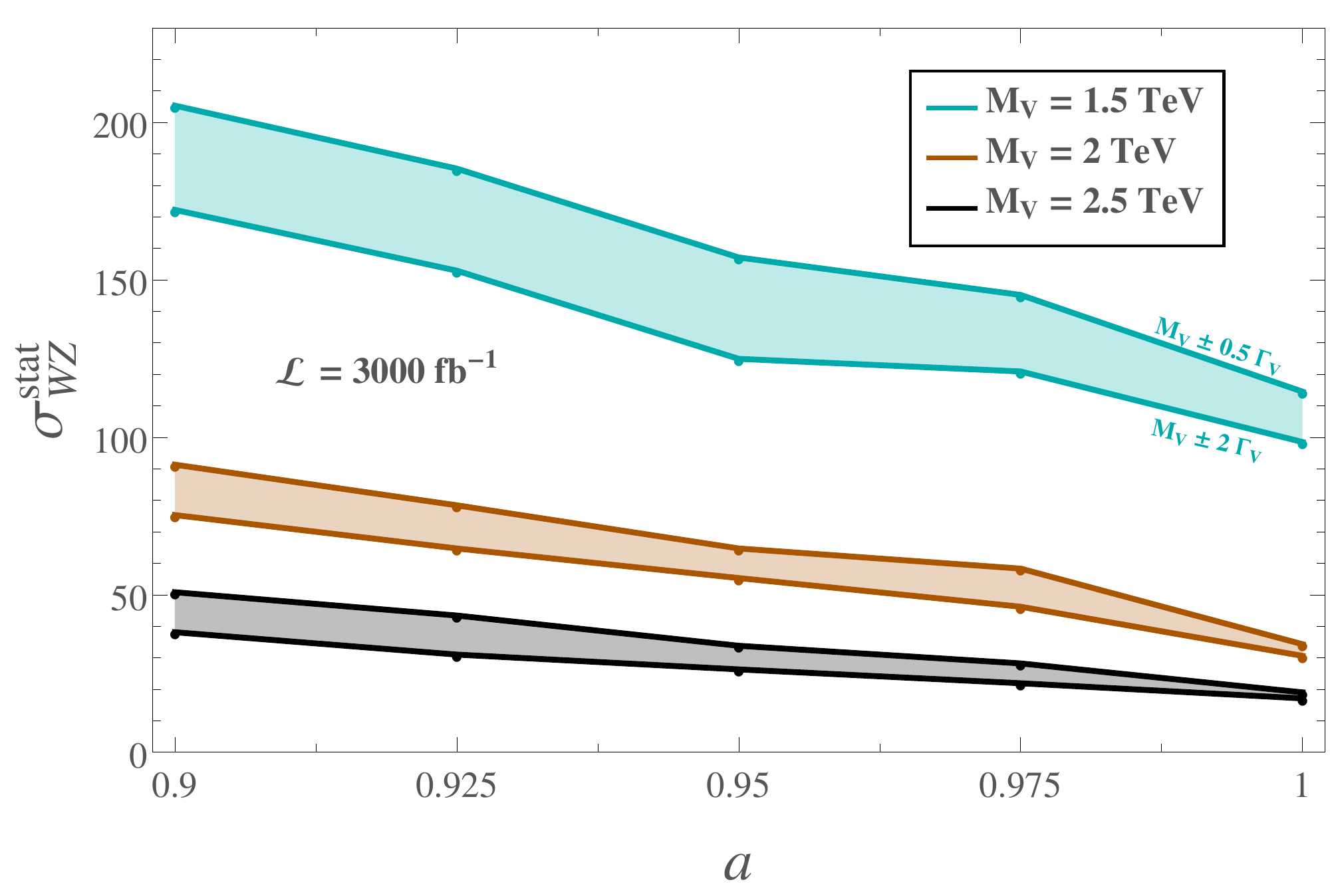}}
\end{minipage}
\hfill
\begin{minipage}{0.32\linewidth}
\centerline{\includegraphics[width=\linewidth]{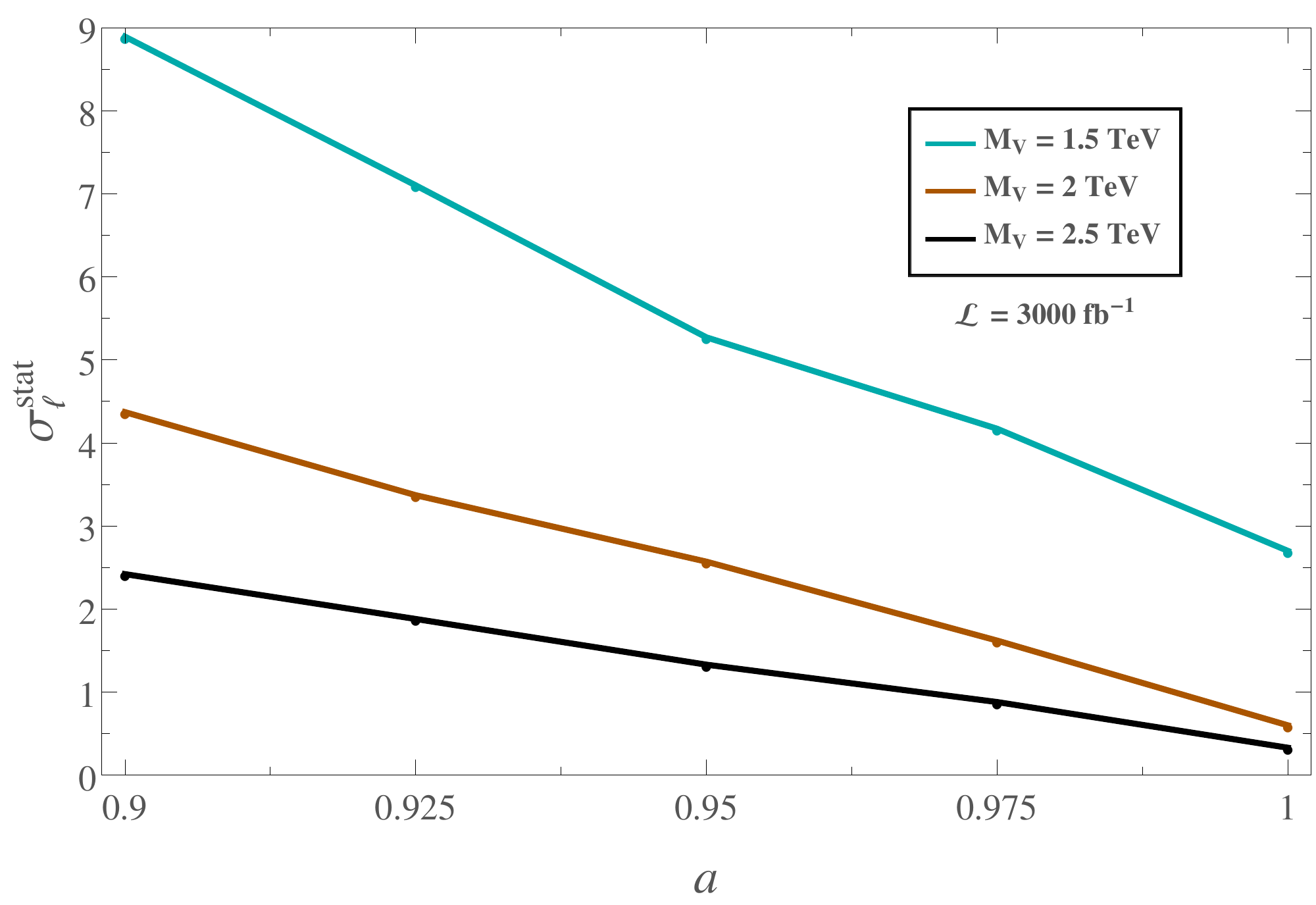}}
\end{minipage}
\hfill
\begin{minipage}{0.32\linewidth}
\centerline{\includegraphics[width=\linewidth]{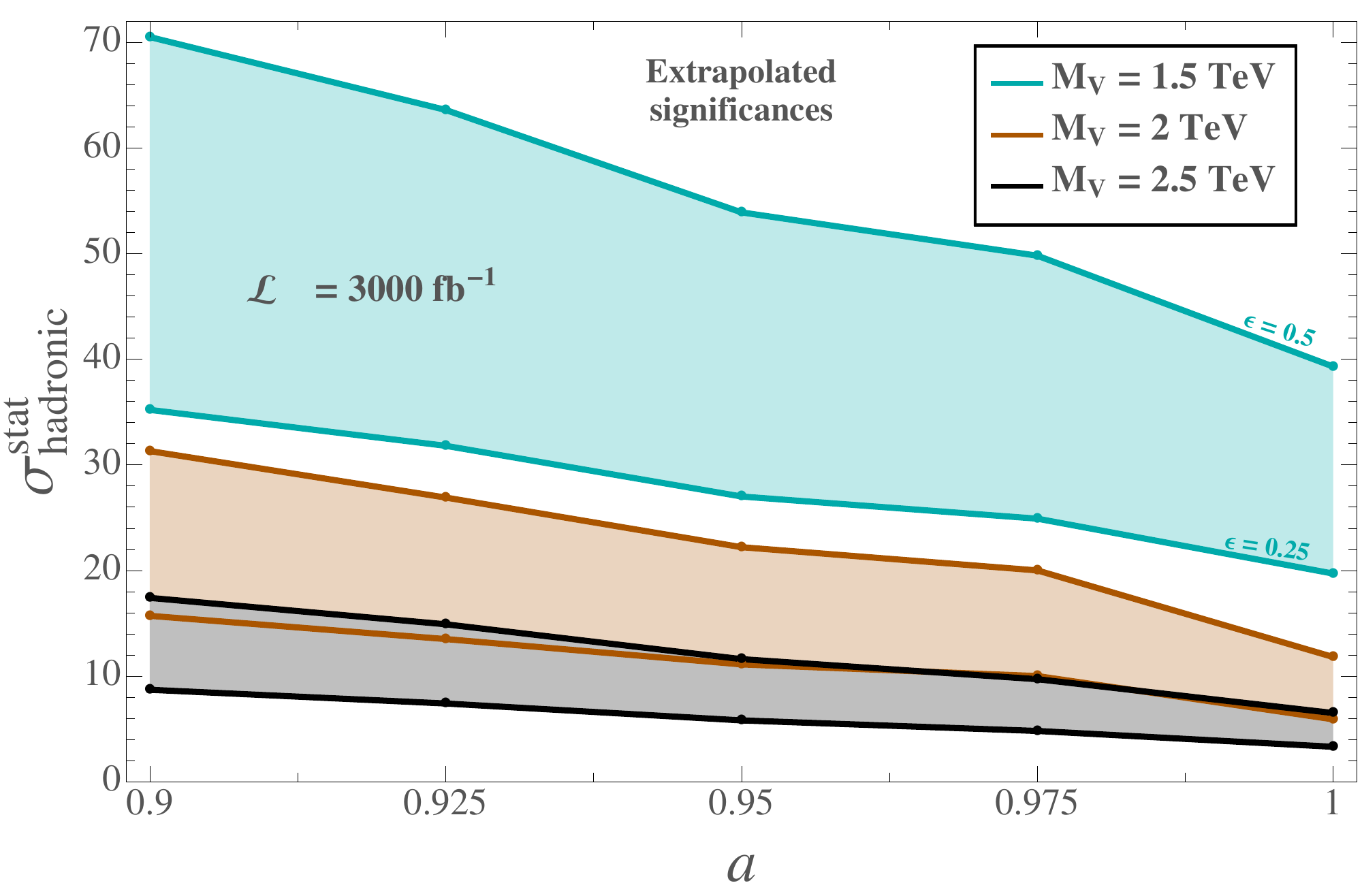}}
\end{minipage}
\caption[]{Predictions for the statistical significance of $W^+Zjj$ with events summed over different intervals (left), $\ell_1^+\ell_1^-\ell_2^+ \nu jj$ optimizing $\sigma^{\rm stat}_{\ell}$(middle) and $JJjj$ with two reconstruction efficiencies (right) as a function of the parameter $a$ for $\mathcal{L}=3000~{\rm fb}^{-1}$.}
\label{fig:significances}
\end{figure}

In Fig.(\ref{fig:significances}) we present the statistical significances for $\mathcal{L}=3000\, {\rm fb}^{-1}$ of our IAM generated vector resonances as a function of the parameter $a$ for the three types of events that we have studied. We can see that if the gauge bosons could be detected, for $\mathcal{L}=300\, {\rm fb}^{-1}$ already, the LHC could be sensitive to the whole allowed interval of $a$. The case with $W,Z$ fat jets in the final state seems very promising, as already for $\mathcal{L}=300\, {\rm fb}^{-1}$ (trivially scaled from Fig.(\ref{fig:significances})) all the mentioned parameter space could be probed through resonances with masses of 1.5 TeV. Values of $a$ that differ from the SM could be tested even relying on the heaviest resonances. The leptonic case requires more luminosity to be observed, but with $\mathcal{L}=3000\, {\rm fb}^{-1}$ the lighter resonances will make possible to explore the entire allowed interval of $a$, and those with masses of 2 TeV could be used to probe deviations from the SM.
\section{Conclusions}
In this work we have explored the production and sensitivity to vector resonances at the LHC in the VBS process $pp \to W^+Zjj$. We have worked under the framework of the EChL supplemented by another effective chiral Lagrangian to describe the vector resonances that are dynamically generated with the use of the IAM and that is suitable for a MonteCarlo analysis.

We have given predictions for the statistical significances of theses resonances in $pp \to W^+Zjj$, $pp \to \ell_1^+\ell_1^-\ell_2^+ \nu jj$ and $pp \to JJjj$ events. We have concluded that, for expected LHC luminosities, the study of these resonances, with masses in the TeV range, would be possible and could let us probe the whole allowed interval for some of the chiral parameters. In particular, the parameter $a$ could be tested at LHC at $\mathcal{L}=300~{\rm fb}^{-1}$ in the $WZ$ final state though resonances with masses between 1.5 and 2.5 TeV and in the $JJjj$ final state through the lightest resonances (although heavier ones would allow to probe values of $a$ that differ from 1). The leptonic case requires higher luminosities to be observed, but for the High Luminosity LHC, resonances with masses of 1.5 TeV will allow to test $a\in[0.9,1]$ and heavier resonances could give us information of whether $a\neq 1$.
\section*{Acknowledgments}
C. G. G. wishes to thank the Moriond organizing committee for its financial support. This work is supported by the ITN ELUSIVES H2020-MSCA-ITN-2015//674896 
and the RISE INVISIBLESPLUS H2020-MSCA-RISE-2015//690575, 
by the Spanish MINECO through the projects FPA2013-46570-C2-1-P, FPA2014-53375-C2-1-P, FPA2016-75654-C2-1-P, FPA2016-76005-C2-1-P,  FPA2016-78645-P
(MINECO/ FEDER, EU), SEV-2012-0249 and SEV-2016-0597, and by the ``Mar\'ia de Maeztu'' Programme under grant MDM-2014-0369.
R.L.D is supported by the ``Ram\'on Areces'' Foundation. 


\section*{References}
\begin{thebibliography}{99}

\bibitem{Appelquist:1980vg}Appelquist, T. {\it et al}, \Journal{{\em Phys. Rev.}}{D22}{200}{1980}.

\bibitem{Longhitano:1980iz}Longhitano, A. C., \Journal{{\em Phys. Rev.}}{D22}{1166}{1980}.

\bibitem{Dobado:1989ax}Dobado, A. {\it et al}, \Journal{{\em Phys. Lett.}}{B228}{495-502}{1989}.

\bibitem{Delgado:2014jda}Delgado, R. L. {\it et al}, \Journal{{\em JHEP}}{07}{149}{2014}.

\bibitem{Alonso:2012px}Alonso, R. {\it et al}, \Journal{{\em Phys. Lett.}}{B722}{330-335}{2013}.

\bibitem{Espriu:2012ih}Espriu, D. {\it et al}, \Journal{{\em Phys. Rev.}}{D87}{055017}{2013}.

\bibitem{Buchalla:2013rka}Buchalla, G. {\it et al}, \Journal{{\em Nucl. Phys.}}{B880}{552-573}{2014}.

\bibitem{Delgado:2017cls}Delgado, R. L. {\it et al}, \Journal{{\em JHEP}}{11}{098}{2017}.

\bibitem{Pich:2015kwa}Pich, A. {\it et al}, \Journal{{\em Phys. Rev.}}{D93}{055041}{2016}.

\bibitem{Ecker:1989yg}Ecker, G. {\it et al}, \Journal{{\em Phys. Lett.}}{B223}{425-432}{1989}.

\end{thebibliography}
\end{document}